\documentclass[11pt]{article}
\usepackage{amssymb,amsthm,amsmath,tikz,pgf,mathtools}
\usepackage[lined,linesnumbered,ruled]{algorithm2e}
\usepackage{epsfig,amsmath, amsfonts,amstext, amsthm, latexsym, graphicx}
\usepackage{epstopdf}
\usepackage{graphicx}

\newtheorem{cor}{Corollary} 
 
\newtheorem{defin}{Definition} \newtheorem{teo}{Theorem}
\newtheorem{lemma}{Lemma}

\begin{document}
\title{Cops and robber on grids and tori}

\author{Fabrizio Luccio and Linda Pagli\\
Universit\`{a} di Pisa, Italy\\ luccio,pagli@di.unipi.it\\
}

\date{}

\maketitle

\begin{abstract} 
This paper is a contribution to the classical cops and robber problem on a graph, directed to two-dimensional grids and toroidal grids.
These studies are generally aimed at determining the minimum number of cops needed to capture the robber and proposing algorithms for the capture.
We apply some new concepts to propose a new solution to the problem on grids that was already solved under a different approach, and apply these concepts to give efficient algorithms for the capture on toroidal grids, that is on semi-tori  (grids with toroidal closure in one dimension) and tori. Then
we treat the problem considering any number $k$ of cops and give efficient algorithms for this case on grids and tori, computing lower and upper bounds on the capture time. Conversely we determine the minimum value of $k$ needed for any given capture time, and study a possible speed-up phenomenon using our algorithms. 

\end{abstract}



\section{Introduction}\label{sec:intro}

The problem of cops and robber on a graph has received considerable attention. Started as a pure pursuit-evasion game it has shown interesting theoretical implications and importance in graph searching, network decontamination, motion planning, security and environment control. As a consequence many versions of the problem have been studied, tipically depending on the type of graph, the knowledge of the actors on the positions of the others, the type and speed of movements allowed. 

In the basic version of the problem cops and robber stay on the vertices of a graph and can move to adjacent vertices or stay still, starting from initial positions chosen first by the cops, then by the robber. The chase proceed in rounds, each of which is composed of a parallel move of the cops followed by a move of the robber which is captured when a cop reaches its vertex and the game terminates. 
We study the problem in this basic version if the graph is a toroidal grid, also revisiting some known results on grids.

\subsection{A brief analysis of the literature}

The cops and robber problem was defined by Quillot~\cite{Q85} and Nowakowski and Winkler~\cite{NW83} as a pursuit-evasion game with one cop, to generate a complex theory in the following years. 
After studying graphs where the game can be won by a single cop, the attention was directed to solve the problem on different classes of graphs with a minimum number of cops, called the {\em cop number}. 
A general survey in this direction can be found in the comprehensive book by Bonato and Nowakowski~\cite{BN11} which brings together the main structural and algorithmic results on the field known when the book appeared. In particular they thoroughly discuss the still open Meyniel conjecture on the sufficiency of $\sqrt{n}$ cops for capturing the robber in an arbitrary graph of $n$ vertices.

Many variants of the basic problem exist for general graphs or for particular classes of graphs, such as considering more than one robber; or cops and robber moving at different speeds; or a robber being invisible for some rounds; or, more recently, the robber escaping surveillance if it maintains a given distance from the cops \cite{C+16}.

With specific reference to two dimensional grids and toroidal grids studied in this paper, the proof that the cop number is 2 in a two-dimensional grid was originally given in \cite{MM87}, and the capture time was determined in \cite{M11}. The cop number 2 for semi-tori and 3 for tori can be derived from the results of \cite{NN98} where the rob capture is studied for products of graphs.
Several variations were proposed, in particular if the visibility of each cop is limited to edges and vertices of its row or column. In \cite{D92, N96} the cops win if they can see the robber, and in \cite{SS89} it is shown that the problem with limited visibility has application in motion planning of multiple robots. 
The study of \cite{SS89} has been revisited in \cite{DKSZ08} and algorithms for the capture using one, two or three cops having constant maximal speed are given. In \cite{B+05} the cop number is determined if the robber can move at arbitrary speed.
A more recent work \cite{EW08} assumes that the initial positions of cops and robber are chosen randomly. In \cite{B+10} the study is extended to $n$-dimensional grids.
As some of the grids studied here are planar, it is worth noting that the problem has been studied for planar graphs in \cite{AF84}, where it is shown that the cop number for such graphs is at most 3,  and in \cite{PX16} where a strategy is presented with capture time $\leq 2$$n$ for graphs of $n$ vertices.

Other important problems with a relation with the one of cops and robber were born in the field of distributed computing with moving agents: such as graph search, see for example \cite{Me+88}; intruder capture, see for example \cite{B+06, INS09}; and network decontamination, see for example \cite{LP09}. A survey of many of these problems can be found in \cite{A04}.

Finally we recall some studies on complexity issues related to the cops and robber problem for general graphs. In \cite{GR95} it was proved that determining the cop number is EXPTIME-complete if the initial positions of cops and robber are given, and in \cite{K15} it has been shown that the problem is EXPTIME-complete without those restrictions. In  fact, in \cite{F+10} it is proved that computing the cop number is NP-hard. Changing the perspective, in \cite{B+06} it was shown that determining the number of cops needed for the capture in no more than a given capture time is NP-hard, and this result was also found in \cite{B+09}.
\subsection{Our contribution}

Among a wealth of possibilities, we limit our treatment to the standard game on 2-dimensional grids starting from the results of \cite{M11}, and then extend it to toroidal grids.  
Our new results are the following. In Section \ref{sec:model} we introduce some new concepts on the capture valid for general graphs, to be used in Section \ref{sec:grids} for showing how the results of \cite{M11} can be found with a new different approach. In Section \ref{sec:tori} this approach is applied to semi-tori (i.e. grids with toroidal closure in one dimension, also called cylindrical grids) 
and to tori, for which we give efficient algorithms for the capture that use two or three cops respectively together with a new proof that these numbers are the minimal possible. 
In Section \ref{sec:k-cops} we then treat the capture problem as a function of any (hence not necessarily minimum) number of cops, giving efficient algorithms also for these cases.

For any given capture time $t^*$ we also determine the minimum number of cops needed for the capture in at most $t^*$ rounds using our algorithms. In this context we adopt the concept of {\em work} $w_k=k\cdot t_k$ of an algorithm run by $k$ cops in total time $t_k$ inherited from parallel processing \cite{KR90,LPP92}, discussing the {\em speed-up} that emerges using a larger number of cops. Note that a related study has been carried out in \cite{LP16} for butterfly decontamination.


\section{Basic model and properties}\label{sec:model}

We adopt the basic model of the cops and robber problem on a simple undirected and connected graph $G=(V,E)$. One or more cops and one robber, collectively called {\em agents}, are placed on the vertices of $G$. The game develops in consecutive rounds, each composed of a cops turn followed by a robber turn. In the cops turn each cop may move to an adjacent vertex or stay still. In the robber turn, the robber may move to an adjacent vertex or stay still. The game is over when a cop reaches the vertex of the robber. 

The initial positions of the cops are arbitrarily chosen, then the initial position of the robber is chosen accordingly. The aim of the cops is capturing the robber in a number of rounds as small as possible, called {\em capture time} $t$; while the robber tries to escape the capture as long as possible. If needed two or more cops can stay on the same vertex and move along the same edge. All agents are aware all the time of the locations of the other agents.
$k$, the {\em cop number}, denotes the smallest number of cops needed to capture the robber. 

As discussed in the following sections we will direct our study to 2-dimensional grids, whose bounding edges may also be connected in the form of a semi-torus or a torus. However, first we pose
some preliminary properties valid for all undirected and connected graphs $G=(V,E)$, partly extending known facts. 
For a vertex $v\in V$, let $N(v)$ denote the set of neighbors of $v$, and let $N[v]=N(v)\cup\{v\}$ denote the closed set of neighbors. We pose:

\begin{defin} \label{def:siege}
A {\em siege} $S(v)$ of a vertex $v$ is a minimum set of vertices containing cops, such that at least one vertex $u\in S(v)$ is in $N(v)$, and $\bigcup_{w\in S(v)}N[w]\supseteq N(v)$. 
Among all the sieges of $v$, $\bar{S}(v)$ denotes one of these sets of minimal cardinality.
\end{defin}

\begin{figure}[h]
\begin{center}
\includegraphics[scale=0.6]{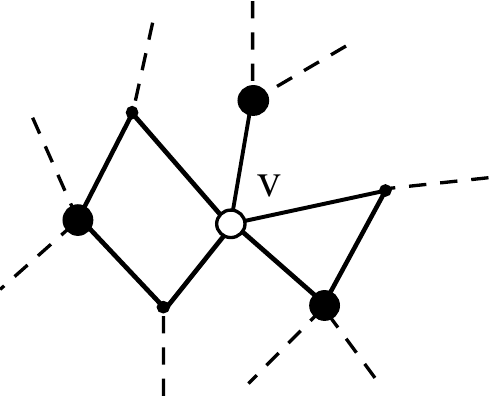}
\end{center}
\caption {\small A minimal siege $\bar{S}(v)$ with the robber (white circle) in $v$ and three cops (black circles) in $\bar{S}(v)$. \label{fig:siege}}\end{figure}

Definition \ref{def:siege} depicts the situation shown in Figure \ref{fig:siege}, where black and white circles on the graph denote vertices occupied by the cops, or by the robber, respectively. Let the robber be in $v$, and assume that the cops have just been moved into the vertices of $S(v)$. Now the robber has to complete the current round, but whether it moves or stands still it will be captured in the next round. In fact the condition $\bigcup_{w\in S(v)}N[w]\supseteq N(v)$ indicates that all the escape routes for the robber have been cut. We immediately have:

\begin{lemma}\label{lem:capture}
The robber is captured in round $i$ if and only if at round $i-1$ the robber is in a vertex $v$ and there is a siege $S(v)$.
\end{lemma}

A lower bound on the number $k$ of cops needed to capture the robber immediately follows from Lemma \ref{lem:capture}, namely:

\begin{lemma}\label{lem:siege}
Let $v$ be a vertex for which $\bar{S}(v)$ has minimal cardinality among all the vertices of the graph. Then $k\geq |\bar{S}(v)|$.
\end{lemma}

Based on the definition of siege we can also establish a lower bound on the capture time $t$. In fact once the initial positions of the cops have been chosen, we shall determine an initial position and a moving strategy of the robber that forces the cops to make at least a certain number of moves for establishing a siege. For this purpose we pose:

\begin{defin} \label{def:siege2}
For a graph $G=(V,E)$ and an integer $e\geq 4$, an e-{\em loop} $L$ is a chordless cycle of e vertices 
where each vertex of \,$V\setminus L$ is adjacent to at most one vertex of $L$.
\end{defin}

Note that a single cop would chase forever a robber that moves inside an $e$-loop. We have:

\begin{lemma}\label{lem:siege3}
Let the initial positions of the $k$ cops $c_1,c_2,\dots, c_k$ be established; let $v$ be the initial position of the robber;
let $d_1\leq d_2\leq \dots \leq d_k$ be the distances (number of edges in the shortest paths) of $c_1,c_2,\dots, c_k$ from $v$; and let $h$ be the cardinality of a minimal siege for $G$, $2\leq h \leq k$. We have:

 \vspace{1mm}
 (i) $t\geq d_1$;
 
 \vspace{1mm}
 (ii) if $v$ belongs to an $e$-loop, 
 $t\geq d_h-\lfloor\frac{e}{2}\rfloor$.

\vspace{2mm}
\noindent {\bf Proof.} {\em (i) Since the robber must be reached by one cop, and may stand still until a cop becomes adjacent to it, we have $t\geq d_1$. 

\noindent (ii) At least $d_1-1$ rounds are needed to bring $c_1$ to a vertex adjacent to $v$, while  the other cops may also move towards the robber. At this point either the robber is surrounded by a siege and the game ends in the next round with $t=d_1$; or the robber completes the current round moving away, in particular to another vertex of the $e$-loop that may be one step closer to the other cops. For the capture, $c_1$ must wait the arrival of other $h-1$ cops to set up a siege and may chase the robber along the $e$-loop to force it to move further towards the other cops up to a loop vertex $v'$. For computing a lower bound, $v'$ must be at a minimum distance from the other cops, and this happens if the robber moves of
$\lfloor\frac{e}{2}\rfloor-1$ further positions after the one reached in the first move when $c_1$ has reached it. 
Since in a siege cop $c_h$ must be at a distance 2 from the robber, the total number of moves of $c_h$ must be at least $d_h-(\lfloor\frac{e}{2}\rfloor-1)-2$, and the final capture move in the siege must be added to this number.}
\hfill $\Box$
\end{lemma}



\section{Capture on grids}\label{sec:grids}

An elegant approach to studying the capture on an $m\times n$ grid has been presented in \cite{M11}, where the grid is treated as the Cartesian product of two paths. This leads to prove that two cops are needed, the capture time is $t=\lfloor\frac{m+n}{2}\rfloor-1$, and this result is optimum. We examine this problem under a different viewpoint, as a basis for studying robber capture on toroidal grids.

Formally an $m\times n$ {\em grid} $G_{m,n}$ is a graph whose vertices are arranged in $m$ rows and $n$ columns, where
each vertex $v_{i,j}$, $0\leq i \leq m-1$ and $0\leq j \leq n-1$
is connected to the four vertices $v_{i-1,j}$, $v_{i+1,j}$, $v_{i,j-1}$, $v_{i,j+1}$, whenever these indices stay inside the closed intervals $[0, m-1]$ and $[0,n-1]$ respectively. Then the vertices can be divided into three sets, namely:
{\em corner vertices}, where both the subscripts $i$ and $j$ have the values 0 or $m-1$, and 0 or $n-1$, respectively; {\em border vertices}, where one of the subscripts $i$ and $j$ has the value 0 or $m-1$, or 0 or $n-1$, respectively; {\em internal vertices}, i.e. all the others. Corner, border, and internal vertices have two, three, and four neighbors each.

\begin{figure}[h]
\begin{center}
\includegraphics[scale=0.6]{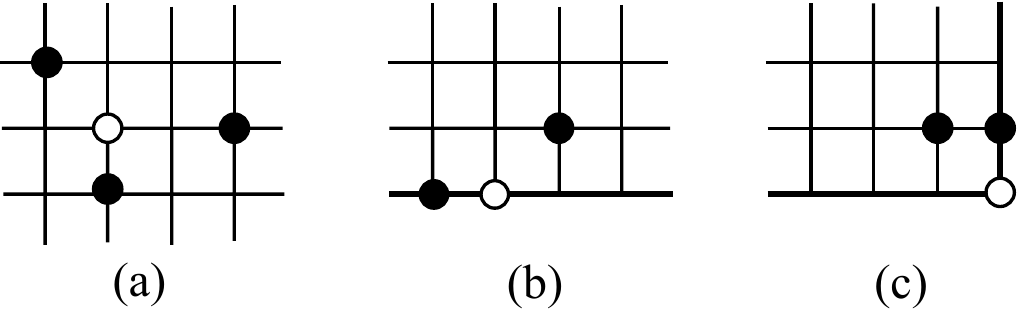}
\end{center}
\caption {\small Examples of a siege $S(u)$ in a grid, if $u$ is an internal vertex, a border vertex, or a corner vertex. \label{fig:sgrid}}
\end{figure}

If two vertices $u,w$ of a grid are adjacent, the set $N(u)\cap N(w)$ is empty. If $w$ is at a distance two from $u$, the set $N(u)\cap N(w)$ contains one or two vertices. This implies that the siege $S(u)$ has cardinality three if $u$ is an internal vertex, or cardinality two if $u$ is a border or corner vertex, see Figure \ref{fig:sgrid}. This bears some initial consequences.

Since the minimal siege for a grid has cardinality two, the number of cops needed to capture the robber is $k\geq 2$ by Lemma \ref{lem:siege}, and in fact two cops suffice as already proved in \cite{M11}. Moreover Lemma \ref{lem:capture} shows that any algorithm using two cops must push the robber to a border or to a corner vertex to establish a siege around it, as three cops would be needed for a siege around an internal vertex. This is what our algorithm will do, capturing the robber in $\lfloor\frac{m+n}{2}\rfloor-1$ rounds as in \cite{M11}. 

Furthermore it can be easily shown that, wherever the cops are initially placed, there is a vertex $v$ where the robber can be placed that is at a distance $d_1\geq \lfloor\frac{m+n}{2}\rfloor-1$ from the closest cop, or at a distance $d_2\geq\lfloor\frac{m+n}{2}\rfloor+1$ from the other cop. Since all the vertices of a grid belong to an $e$-loop consisting of square cycles of $e=4$ vertices we have 
$\lfloor\frac{e}{2}\rfloor =2$, hence $t\geq \lfloor\frac{m+n}{2}\rfloor-1$ by Lemma \ref{lem:siege3} case (i) or (ii), that confirms the lower bound of \cite{M11}.

\begin{figure}[h]
\begin{center}
\includegraphics[scale=0.6]{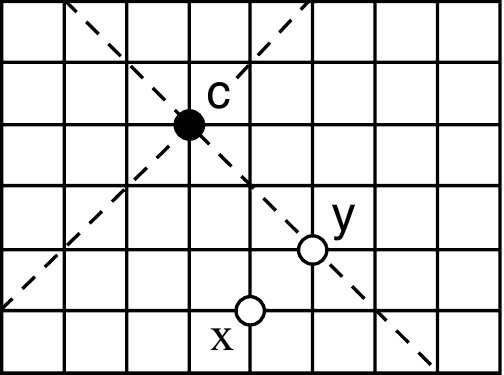}
\end{center}
\caption {\small A cop $c$ divides the grid in four zones limited by two straight lines at $\pm 45^{\circ}$ passing through the vertex of $c$, and by the grid border. Two positions $x,y$ of the robber are shown, inducing different cop movements. \label{fig:cone}}
\end{figure}

A new concept is the one of the {\em shadow cone} of a cop $c$, namely a zone (set of vertices) of the grid ending on the border, from where the robber is impeded by $c$ to escape. To this end let $c$ be in vertex $u=v_{i,j}$ and consider two straight lines at $\pm 45^{\circ}$ through $u$  that divide the grid into four zones whose borders contain vertices placed on the two lines, called {\em edges} of the zone, and vertices placed on the border of the grid, see Figure \ref{fig:cone}. The shadow cone of $c$ is one of the four zones, chosen by $c$. In particular the robber is said to stay {\em within} the cone if it stays {\em in} the cone but not on one of its edges. 

Without loss of generality, let the shadow cone of $c$ lay ''below'' the cop as shown in Figure \ref{fig:cone}.  Two cases may occur, to which the following Cone Rule applies. 
Once the shadow cone is established at the beginning of the operations as the one containing the robber, the role of the Cone Rule is to keep the robber in the cone, possibly moving the cone by one position to compensate the robber's movement (point 2.$iv$ of the rule). An important consequence is the following lemma.


\begin{center}
\fbox{
\begin{minipage}{12cm}
\small{

\vspace{2mm}
\noindent{\bf CONE RULE}

\vspace{2mm}
\hspace{4mm}The cop $c$ is in vertex $v_{i,j}$, and the robber is in the shadow cone of $c$.

\begin{enumerate}

\item Let be the cop's turn to move. $(i)$ If the robber is within the cone (vertex $x$ in Figure \ref{fig:cone}) the cop moves ``down'' to vertex $v_{i+1,j}$ thereby reducing the size of the cone while keeping the robber in it. $(ii)$ If  the robber is on an edge of the cone (e.g. in vertex $y$) the cop does not move. 

\item Let be the robber's turn to move. $(iii)$ If the robber remains in the cone the subsequent cop's move takes place as specified in points 1.$i$ or 1.$ii$ whichever applies. $(iv)$ If the robber moves out of the cone from one of its edges (e.g. from vertex $y$, moving ``up'' or ``to the right''), the cop moves across to vertex $v_{i,j+1}$ thereby shifting its shadow cone by one positions to keep the robber in the cone.

\vspace{2mm}

\end{enumerate}
}
\end{minipage}
}
\end{center}


\begin{lemma}
\label{lemma:cone} By applying the Cone Rule, if the robber reaches an edge of the cone it will never be able to reach the opposite edge. 

\vspace{1mm}
\noindent {\bf Proof.} {\em For reaching the opposite edge of the cone the robber should traverse the cone, and at each round the cop would become closer to it (point 2.$iii$ of the cone rule), to become adjacent to the robber in the center of the cone and capture it.}
\hfill $\Box$
\end{lemma}


Note that if at each round the robber stands on an edge of the cone, the cop would not be able to reach the robber. In fact two cops $c_1$, $c_2$ are needed for the capture. We now give the following algorithm GRID that runs in $t=\lfloor\frac{m+n}{2}\rfloor -1$ rounds as for the algorithm of \cite{M11}, but  is useful for the discussion that follows on toroidal grids.
W.l.o.g. we let $m\leq n$, $m_1=\lfloor\frac{m-1}{2}\rfloor$, $m_2=\lceil\frac{m-1}{2}\rceil$, $n_1=\lfloor\frac{n-1}{2}\rfloor$, $n_2=\lceil\frac{n-1}{2}\rceil$. 

\begin{figure}[h]
\begin{center}
\includegraphics[scale=0.6]{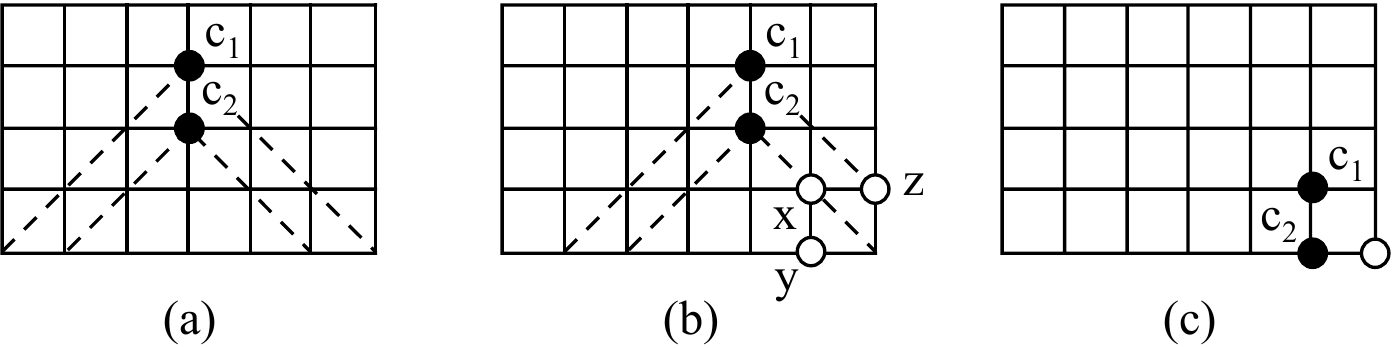}
\end{center}
\caption {\small (a) Initial placement of the two cops for grids with $m$ odd and $n$ odd, denoted as $o\,|\,o$. \,(b) The cops push the robber towards the border of the grid. A similar situation occurs for grids $e\,|\,o$ and  $o\,|\,e$, and for grids $e\,|\,e$ exchanging rows with columns. Vertices $x,y,z$ indicate particular positions of the robber. Vertices $x,y$ are in the two cones as in step 4.1 of the algorithm GRID. Vertex $z$ is in one of the two possible conditions indicated in step 4.2 of the algorithm. \,(c) The final siege.
\label{fig:2cgrid}}
\end{figure}

\vspace{2mm}
To understand how the algorithm works some observations are in order.
The cops are initially adjacent, then their cones have a large portion in common but their edges are disjoint, see Figure \ref{fig:2cgrid}.
Then, up to the round in which the siege is established, they move in parallel so their mutual positions do not change. 
At the beginning the robber is in at least one of the shadow cones and is kept in this condition after each cops' move. 
Note that to delay the capture as much as possible the robber must eventually escape from one side of a cone. By Lemma \ref{lemma:cone}, however, it is forced to escape always from the same side until it ends up in a siege, in a grid corner. We can state the following Theorem \ref{teo:grid}.


\vspace{2mm}
\begin{center}
\fbox{
\begin{minipage}{11.5cm}
\small

\vspace{1mm}
\noindent {\bf algorithm} GRID($m$,$n$)

\vspace{2mm}

1.\; initial positions of the cops $c_1,c_2$: 

\hspace{6mm}for $m$ even and $n$ odd ($e\,|\,o$), or for  $e\,|\,e$, {\bf place}
$c_1$ in $v_{m_1,n_1}$ and $c_2$ 

\hspace{6mm}in $v_{m_2,n_1}$; 
for $o\,|\,o$, {\bf place} $c_1$ is in $v_{m_1-1,n_1}$ and $c_2$ is in $v_{m_1,n_1}$, see 

\hspace{6mm}figure \ref{fig:2cgrid}.$a$; 
for $o\,|\,e$, {\bf place} $c_1$ is in $v_{m_1,n_1}$ and $c_2$ is in $v_{m_1,n_2}$; 

// \hspace{1mm}assume to work on an $o\,|\,o$ grid (the others are treated similarly)


// \hspace{1mm}the shadow cones are chosen so that at least one of them will contain

\hspace{5mm}  the robber; assume that they lay below the cops as in Figure \ref{fig:2cgrid}

\vspace{2mm}
2.\; initial position of the robber: 

\hspace{6mm}{\bf place} the robber in any vertex not adjacent to a cop\,;



\vspace{2mm}
3.\; {\bf repeat} 


\vspace{1mm}
3.1\; \hspace{4mm}
{\bf if} (the robber is in the two cones)   

\vspace{1mm}
\hspace{20mm}{\bf move} both cops one step down 

\vspace{1mm}
3.2\; \hspace{4mm}
{\bf else} (the robber is on the edge of a cone but outside the other 

\hspace{20mm}cone) {\bf or} (the robber is outside the two cones) 


\vspace{1mm}
\hspace{20mm}{\bf move} both cops horizontally in the direction of the robber\,;



\vspace{1mm}
\hspace{12mm}{\bf move} the robber in any way to try to escape from the cops

\vspace{2mm}
4.\; {\bf until} the robber makes its last move inside a siege;

// \hspace{1mm} the siege is established with the robber in a grid corner (figure \ref{fig:2cgrid}.c)

\vspace{2mm}
5.\; {\bf capture} the robber;

\vspace{2mm}
\end{minipage}
}

\end{center}


\vspace{3mm}

\vspace{3mm}

\begin{teo}
\label{teo:grid}
In a grid $G_{m,n}$ two cops can capture the robber in $t=\lfloor\frac{m+n}{2}\rfloor -1$ rounds. 

\vspace{1mm}
 \noindent {\bf Proof.}  {\em Use algorithm GRID. }
 
 \noindent(i) Capture. {\em After the execution of one round (steps 3, 4), either the shadow cones become tighter around the robber (step 3.1), or the robber exits from the edge of one cone and moves towards the border of the grid. By Lemma \ref{lemma:cone} this exit may be repeated in further rounds only from the same edge. Then the robber will be eventually pushed by the cops to the border of the grid, and from there to a corner, surrounded by a siege. Step 5 completes the capture. }

\vspace{1mm}
\noindent (ii) Evaluation of $t$. {\em Each movement of the cops up to establishing a siege reduces by one their distance from a grid corner where the robber is eventually pushed. With the chosen initial displacement, $d=\lfloor\frac{m+n}{2}\rfloor -1$ is the maximum distance between a grid corner and the closest cop. Then the siege is  reached in $d-1$ rounds and the capture is done in $d$ rounds.}
\hfill $\Box$
\end{teo}

As already noted the capture time given in Theorem \ref{teo:grid} is equal to the one proved in \cite{M11}.


\section{Capture on toroidal grids}\label{sec:tori}

We now extend our study to 2-dimensional grids in the form of {\em semi-tori} $S_{m,n}$ and {\em tori} $T_{m,n}$. In $S_{m,n}$ (also called {\em cylinder}) a toroidal closure occurs in the first dimension, that is each vertex $v_{i,0}$ is connected with $v_{i,n-1}$ by an edge, $0\leq i\leq m-1$. In $T_{m,n}$ the toroidal closure occurs in both dimensions, that is also each vertex $v_{0,j}$ is connected with $v_{m-1,j}$, $0\leq j\leq n-1$. Then semi-tori have a top and a bottom border and tori have no borders.

From the known results on the cop number for the capture on products of graphs proved in  \cite{NN98} we have that 2 and 3 cops are needed for semi-tori and for tori, respectively. Based on Lemma \ref{lem:siege} we confirm these numbers as lower bounds, give two algorithms for the capture that use 2 and 3 cops respectively, and compare the time required by these algorithms with the lower bound given in Lemma \ref{lem:siege3}. 

\subsection{Capture on semi-tori}\label{sec:semi}

Consider a $S_{m,n}$ with the toroidal closures on the rows, and let $m\geq 3, n\geq 4$ to avoid trivial cases. At least two cops are needed since the vertices on the border have a siege of minimum cardinality 2. In fact the following algorithm SGRID uses two cops. 

\begin{figure}[h]
\begin{center}
\includegraphics[scale=0.55]{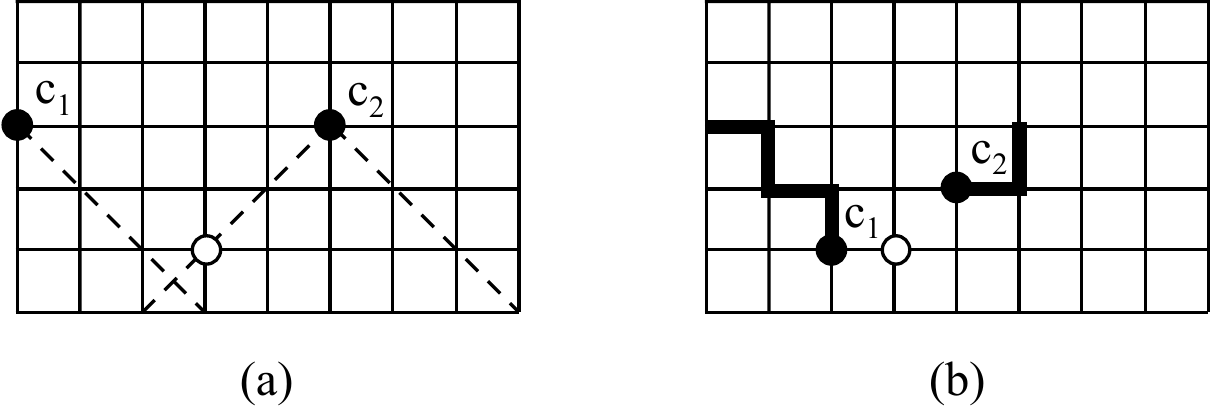}
\end{center}
\caption {\small (a) Initial positions of $c_1,c_2$ in $S_{6,9}$, with $\lfloor\frac{m-1}{2} \rfloor=2, \lceil\frac{n}{2} \rceil=5$. Note the cop-free gaps of 4 and 3 columns between the cops. (b)  Cop paths to a pre-siege.
\label{fig:2sgrid}}
\end{figure}

The cops $c_1,c_2$ are initially placed in row $\lfloor\frac{m-1}{2} \rfloor$, and columns 0 and $\lceil\frac{n}{2} \rceil$ respectively, so that there are two cop-free gaps
of $\lceil\frac{n}{2} \rceil-1$ columns between $c_1$ and $c_2$, and of $\lfloor\frac{n}{2} \rfloor-1$ columns between $c_2$ and $c_1$ (around the semi-torus), see Figure \ref{fig:2sgrid}.a. A crucial configuration is the one of a {\em pre-siege} where 
the robber is in vertex $v_{i,j}$, cop $c_1$ is in vertex $v_{i,j-1}$, and cop $c_2$ is in vertex $v_{i-1,j+1}$, see Figure \ref{fig:2sgrid}.b, so the robber can move only one vertex down and will eventually be pushed to a siege in the bottom row (if the robber is already in the bottom row the pre-siege is in fact a siege).



\vspace{2mm}
\begin{center}
\fbox{
\begin{minipage}{12.2cm}
\small

\vspace{1mm}
\noindent {\bf algorithm} SGRID($m$,$n$)

\vspace{2mm}


\vspace{2mm}
1.\; initial positions of the cops $c_1,c_2$: 

\vspace{1mm}
\hspace{5.5mm}{\bf place}
$c_1$ in $v_{\lfloor\frac{m-1}{2} \rfloor,0}$; {\bf place} $c_2$ 
in $v_{\lfloor\frac{m-1}{2} \rfloor,\lceil\frac{n}{2} \rceil}$; 

\vspace{1mm}
\hspace{5.5mm}{\bf let} $\gamma_1, \gamma_2$ be the shadow cones of $c_1,c_2$;

// \hspace{1mm}assume that $\gamma_1,\gamma_2$ are chosen below the cops as in Figure \ref{fig:2sgrid}.a

\vspace{2mm}
2.\;\; initial position of the robber $r$: 

\vspace{1mm}
\hspace{6mm}{\bf place} the robber in any vertex not adjacent to a cop;


\vspace{2mm}
3.\;\; {\bf repeat} 


\vspace{1mm}
3.1\; \hspace{4mm}
{\bf if} ($r$ is outside $\gamma_1$ and $\gamma_2$)   
{\bf move} $c_1$ and $c_2$ horizontally towards $r$

\vspace{1mm}
3.2\; \hspace{4mm}
{\bf else} 
{\bf if} ($r$ is within $\gamma_1$ and/or within $\gamma_2$)   
{\bf move} $c_1$ and $c_2$ down

\vspace{1mm}
3.3\; \hspace{13mm}{\bf else} 
{\bf if} ($r$ is on an edge of $\gamma_1$ (resp. $\gamma_2$) 

\hspace{34mm}{\bf and} outside $\gamma_2$ (rep. $\gamma_1$))  

\hspace{34mm}{\bf move} $c_2$ (resp. $c_1$) horizontally towards that edge 

\vspace{1mm}
3.4\; \hspace{22mm}{\bf else} 
{\bf if} ($r$ is on an edge of $\gamma_1$ and on an edge of $\gamma_2$)  

\hspace{41mm}\{{\bf if} (the cops are in different rows)

\hspace{50mm}{\bf move} down the cop in the highest row

\hspace{43mm}{\bf else} 
{\bf move} down one of the cops\}; 


\vspace{1mm}
3.5\; \hspace{4mm}{\bf move} the robber in any way to try to escape from the cones

\vspace{2mm}
4.\;\; {\bf until} the robber makes its last move inside a siege;

// \hspace{1mm}the siege is established with the robber on the lower border of the grid

\vspace{2mm}
5.\;\; {\bf capture} the robber;

\vspace{2mm}
\end{minipage}
}
\end{center}


\vspace{2mm}
To understand how algorithm SGRID works observe the following. 

\begin{itemize}

\item If the robber $r$ is initially in a gap between the cops $c_1,c_2$ and outside both cones it will always remain in that gap. In fact the cops move towards $r$ (step 3.1) until $r$ is on the edge of a cone and will not be allowed to reach the opposite edge by the Cone Rule that is enforced in steps 3.3 and 3.4. Therefore in the longest chase the robber will be captured on the bottom border, in a column of the larger gap. 

\item If $r$ is initially within a cone of a cop $c$ it will be kept in this cone by $c$ that moves down reducing the size of the cone around $r$ at each round (step 3.2). If $r$ reaches an edge of the cone, steps 3.3 and 3.4 maintain $r$ in the gap where the capture will take place. 

\item If $r$ is on an edge of the cone of $c$ and not on an edge of the other cone, and $r$ escapes from the cone in its turn to move, then $c$ moves the cone to recapture $r$ in its cone (step 3.1 or 3.3) and the gap between the cops becomes smaller. If $r$ is on an edge of both cones and escapes from one or both of them, it is recaptured in one or both cones (step 3.1 or 3.3).

\item The cops are kept in the same row in steps 3.1, 3.2, and 3.3, and may occupy two adjacent rows in step 3.4. However they will never be at a larger vertical distance, and will regain the same row with a new application of step 3.4. 

\item The cops always move towards $r$, but only one cop moves in a round in steps 3.3 and 3.4.
This has an impact on the capture time that is maximized if the robber forces the cops to repeat these two steps as many times as possible. 

\end{itemize}


\begin{teo}
\label{teo:sgrid}

In a semi-torus $S_{m,n}$  two cops can capture the robber in time:

\vspace{1mm}
{\em (i)}\;\, $t=\lceil\frac{n}{2} \rceil+ 2\lfloor\frac{m}{2}\rfloor -2$,\; for $\lfloor\frac{m}{2}\rfloor \leq \lceil\frac{n-2}{4} \rceil $;
 
\vspace{1mm}
{\em (ii)}\, $t=\lceil\frac{n}{2} \rceil+  \lceil\frac{n-2}{4} \rceil + \lfloor\frac{m}{2}\rfloor -2$,\; for $\lfloor\frac{m}{2}\rfloor > \lceil\frac{n-2}{4} \rceil $.

\vspace{2mm}
\noindent {\bf Proof.} {\em Use algorithm SGRID.}

\noindent Capture. {\em If the robber is within or outside both cones, the two cops become closer to it with one move, and repeat the round until the robber ends up on the edge of one or both cones. Now one cop moves closer to the robber and the other remains still. This inevitably bring cops and robber in a siege  or in a pre-siege  condition, with the robber pushed down to a siege on the border. Then the capture takes place in the next round.}

\vspace{1mm}
\noindent Evaluation of $t$. {\em The cops always move in the direction of the robber, either horizontally or vertically, until the robber is captured in a border vertex $v_{m-1,j}$, with $1\leq j \leq \lceil\frac{n}{2} \rceil-1$. Let $h=\lceil\frac{n}{2} \rceil$ and $k=2\lfloor\frac{m}{2}\rfloor$ respectively denote the sum of the horizontal and of the vertical distances between the initial positions of the two cops $c_1,c_2$ and $v_{m-1,j}$. Note that $h$ and $k$ are independent of $j$. Letting $r_1,r_2,r_3,r_4$ be the number of rounds respectively executed in steps $4.1,4.2,4.3,4.4$ of the algorithm, the siege is reached in $t=r_1+r_2+r_3+r_4$ rounds subject to the conditions $2r_1+r_3=h-2$ and $2r_2+r_4=k-1$, because when the siege is reached the values of $h$ and $k$ are respectively reduced to 2 and 1. The highest value of $t$ then occurs when $r_1$ and $r_2$ are minimized. For this purpose the robber must stay on the edge of a cone and not within the other cone, for as many rounds as possible.

(i) For $\lfloor\frac{m}{2}\rfloor \leq \lceil\frac{n-2}{4} \rceil $ the robber can force the cops to always apply steps 4.3 and 4.4 if it starts in the lowest vertex of the edge of one of the cones, e.g in vertex $v_{m-1,\lfloor\frac{m}{2}\rfloor}$, and stays still until the siege is built. The total number of rounds will be then  $t=r_3+r_4+1$ including the last round following the siege, subject to the conditions $r_3=h-2=\lceil\frac{n}{2} \rceil-2$ and $r_4=k-1= 2\lfloor\frac{m}{2}\rfloor-1$, for a total of $t=\lceil\frac{n}{2} \rceil+ 2\lfloor\frac{m}{2}\rfloor -2$.

(ii) For $\lfloor\frac{m}{2}\rfloor >\lceil\frac{n-2}{4} \rceil$ the two cones $\gamma_1$ and $\gamma_2$ intersect and the cops will always be able to apply steps 4.1 and/or 4.2 for some times. To reduce $r_1$ and/or $r_2$ as much as possible the robber must start on the lowest vertex on the edge of a cone (say $\gamma_1$) and not within the other cone, and stay there until a pre-siege is built. For such a vertex $v_{i,j}$ we have $i=\lfloor\frac{m-1}{2}\rfloor + \lceil\frac{n-2}{4} \rceil$, $j=\lceil\frac{n-2}{4} \rceil$, with $1\leq \lceil\frac{n-2}{4} \rceil\leq \lfloor\frac{m}{2}\rfloor-1$.
To build a pre-siege steps 4.3 and 4.4 are applied in $r_3+r_4$ rounds, subject to the conditions $r_3=h-2=\lceil\frac{n}{2} \rceil-2$, and $r_4=k-1=2\lceil\frac{n-2}{4} \rceil -1$ since $v_{i,j}$ is at a vertical distance $\lceil\frac{n-2}{4} \rceil$ from both cops. We then have $r_3+r_4=\lceil\frac{n}{2} \rceil+2\lceil\frac{n-2}{4} \rceil -3$. Once a pre-siege is built, the robber must move down and step 4.2 is applied until $c_1$ reaches the siege in border in row $m-1$, in $r_2= m-1 - (\lfloor\frac{m-1}{2}\rfloor + \lceil\frac{n-2}{4} \rceil) =  \lfloor\frac{m}{2}\rfloor -\lceil\frac{n-2}{4} \rceil$ rounds. In total $t=r_2+r_3+r_4+1 = \lceil\frac{n}{2} \rceil+ \lceil\frac{n-2}{4} \rceil  + \lfloor\frac{m}{2}\rfloor -2$.
}
\hfill $\Box$
\end{teo}


For $S_{6,9}$ of Figure \ref{fig:2sgrid} we have $\lfloor\frac{m}{2}\rfloor =3$ and $\lceil\frac{n-2}{4} \rceil =2$, so case (ii) of Theorem \ref{teo:sgrid} applies and the capture takes $t_2=5+2+3-2=8$ rounds. 

\vspace{1mm}Since all the vertices of a grid belong to an $e$-loop consisting of square cycles of $e=4$ vertices, a lower bound can be established by Lemma \ref{lem:siege3} on the capture time on $S_{m,n}$. 
We have:

\begin{lemma}\label{lem:semitori}
The capture time in a semi-torus $S_{m,n}$ admits a lower bound $t_{LS}=\lfloor\frac{n}{2} \rfloor+\lfloor\frac{m}{2}\rfloor-2$.

\vspace{2mm}
\noindent {\bf Proof.} {\em For any vertex $u$ of the semi-torus there is a vertex $w$ whose distance from $u$ is at least $\lfloor\frac{n}{2} \rfloor+\lfloor\frac{m}{2}\rfloor$ (e.g this occurs between vertices $v_{0,0}$ and $v_{\lfloor\frac{m}{2}\rfloor,\lfloor\frac{n}{2} \rfloor}$). If $c_2$ is initially placed in $u$, then $r$ can be placed in $w$ and we have from Lemma \ref{lem:siege3}:
$t_{LS}=d_2-\lfloor\frac{e}{2}\rfloor =\lfloor\frac{n}{2} \rfloor+\lfloor\frac{m}{2}\rfloor-2$.
}
\hfill $\Box$
\end{lemma}

Letting $t_{US}$ be the upper bound to $t$ given in Theorem \ref{teo:sgrid} we have:

\begin{cor}
\label{cor:sLB}
In a semi-torus $S_{m,n}$ the ratio $t_{US}/t_{LS}\rightarrow 1$ for $n/m\rightarrow\infty$ and for $n/m\rightarrow 0$.

\vspace{2mm}
\noindent {\bf Proof.} {\em For $n/m\rightarrow\infty$ case (i) of Theorem \ref{teo:sgrid} applies and we have:
$t_{US}/t_{LS}=(\lceil\frac{n}{2} \rceil + 2\lfloor\frac{m}{2}\rfloor  -2)/(\lfloor\frac{n}{2} \rfloor+\lfloor\frac{m}{2}\rfloor-2)\rightarrow 1$.

\noindent For $n/m\rightarrow 0$ case (ii) of Theorem \ref{teo:sgrid}  applies and we have:
$t_{US}/t_{LS}=(\lceil\frac{n}{2} \rceil + \lceil\frac{n-2}{4} \rceil + \lfloor\frac{m}{2}\rfloor  -2)/(\lfloor\frac{n}{2} \rfloor+\lfloor\frac{m}{2}\rfloor-2)\rightarrow 1$.
}

\end{cor}

Corollary \ref{cor:sLB} shows that if $n$ is much greater or is much smaller than $m$, algorithm SGRID tends to be optimal with regard to the capture time.


\subsection{Capture on tori}

The capture on tori $T_{m,n}$ is more difficult as there are no borders where to push the robber. All the vertices now admit a siege of cardinality 3, then at least three cops are needed, see Figure \ref{fig:sgrid}. 
The following capture algorithm TGRID calls the procedures GUARD and CHASE and uses three cops $c_1,c_2,c_3$ with shadow cones $\gamma_1,\gamma_2, \gamma_3$. 

\begin{figure}
\begin{center}
\includegraphics[scale=0.56]{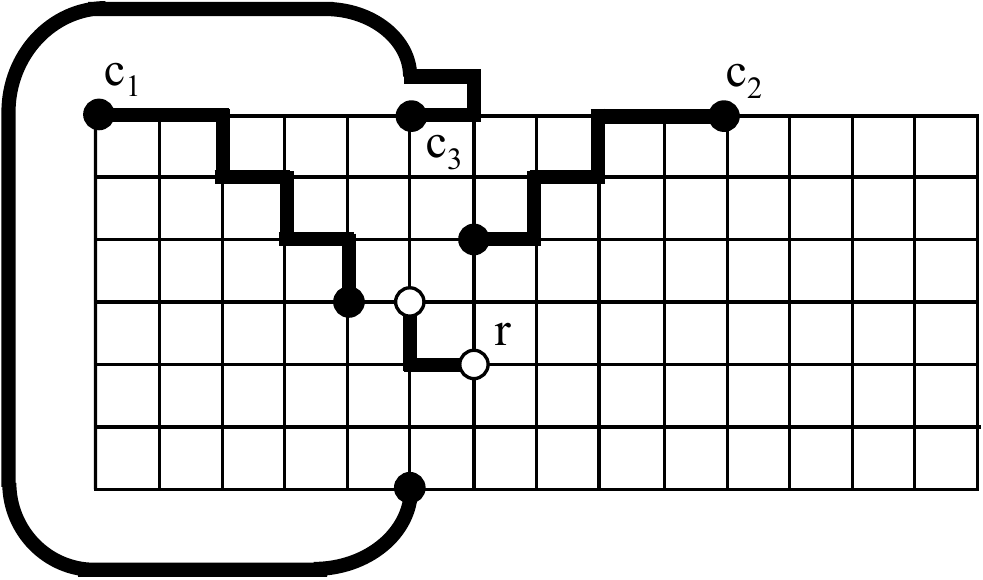}
\end{center}
\caption {\small 
Chase with three cops in $T_{7,15}$ up to a pre-siege, assuming that $c_3$ becomes the guard. The first two moves of $c_2,c_3$, and $r$ take place in the GUARD phase, that ends when $c_2$ reaches column $\lceil\frac{n}{2} \rceil = 8$.
\label{fig:tgrid}}
\end{figure}

Without loss of generality we define the algorithm for $n\geq m$ (simply exchange rows with columns if $m>n$), and let $m\geq 6$ and $n\geq 6$ to avoid trivial cases.  Place all the cops $c_1,c_2,c_3$ in row $0$, and in columns 0, $\lceil\frac{2n}{3} \rceil$, and $\lceil\frac{n}{3} \rceil$, respectively (see Figure \ref{fig:tgrid}). Note that initially there is a cop-free gap of $\lceil\frac{n-3}{3} \rceil$ columns between $c_1$ and $c_3$, and a cop-free gap of $\lceil\frac{n-3}{3} \rceil$ or $\lfloor\frac{n-3}{3} \rfloor$ columns between $c_3$ and $c_2$ and between $c_2$ and $c_1$ around the torus.
Starting with the cops in any row will be the same because we work on a torus. 
The strategy is to bring a cop to {\em guard} the robber $r$ (procedure GUARD), that is the cop will reach the column of $r$ and then follow $r$ if it moves horizontally, so to build a {\em virtual border} along the row of the guard that prevents $r$ from traversing it. When the guard is established, the other cops start chasing $r$ (procedure CHASE) with an immediate extension of algorithm SGRID. Without loss of generality we assume that the initial position of the robber is such that $c_2$ or $c_3$ becomes the guard.


\begin{center}
\fbox{
\begin{minipage}{12.2cm}
\small

\vspace{2mm}
\noindent {\bf algorithm} TGRID($m$,$n$)

\vspace{2mm}
1.\; initial positions of the cops $c_1,c_2,c_3$: 

\hspace{5.5mm}{\bf place}
$c_1$ in $v_{0,0}$; {\bf place} $c_2$ 
in $v_{0,\lceil\frac{2n}{3} \rceil}$; 
{\bf place} $c_3$ 
in $v_{0,\lceil\frac{n}{3} \rceil}$; 

\hspace{5.5mm}{\bf let} $\gamma_1, \gamma_2, \gamma_3$ be the shadow cones of $c_1,c_2,c_3$;

\vspace{2mm}
2.\; initial position of the robber $r$: 

\hspace{5.5mm}{\bf place} $r$ in any vertex not adjacent to a cop;

\vspace{1mm}
//\hspace{1mm} w.l.o.g let the column of $r$ lie in the closed interval 
$[\lceil\frac{n}{3} \rceil : \lceil\frac{2n}{3} \rceil$-$1]$


\vspace{3mm}
3.\; GUARD\,;

\vspace{1mm}
//\hspace{1mm} $c_2$ and $c_3$ move to establish the guard;  upon exit  $c_g$ is the guard 

//\hspace{1mm} 
and $c_h$ is  in column $\lceil\frac{n}{2} \rceil$ to start  chasing $r$ together with $c_1$, 

//\hspace{1mm}
with $g$=2, $h$=3, or $g$=3, $h$=2

\vspace{3mm}
4.\; CHASE\,;

\vspace{1mm}
//\hspace{1mm} $r$ is captured by $c_1,c_h$  with an extension of algorithm SGRID

//\hspace{1mm} $c_g$ is the guard

\vspace{2mm}
\end{minipage}
}
\end{center}


To understand how algorithm TGRID works observe the following.

\begin{itemize}

\item After steps 1 and 2 to establish the initial positions of cops and robber, the algorithm is divided in a phase GUARD to establish the guard $c_g$, with $g$=2 or $g$=3, followed by a phase CHASE of chasing. GUARD is repeated until $c_h$ reaches the column $\lceil\frac{n}{2}\rceil$ to start the chase together with $c_1$ even if the guard has been established in a previous round.


\begin{center}
\fbox{
\begin{minipage}{12.2cm}
\small

\vspace{2mm}
{\bf phase} GUARD

\vspace{2mm}
1.\;\; {\bf let} $y_0,y_1,y_2,y_3$ be the columns of $r,c_1,c_2,c_3$ respectively;

\vspace{0.5mm}
\hspace{5mm} $g=0$;

\vspace{1mm}
//\hspace{1.5mm} $g$=0, $g$=2, $g$=3 respectively denote that: the guard has not 

//\hspace{1.5mm} yet been established, or $c_2$ is the guard, or $c_3$ is the guard;

\vspace{2mm}
2.\;\; {\bf repeat} \hspace{6mm} // establishing the guard


\vspace{1mm}
2.1\; \hspace{4mm} 
{\bf if} ($y_2==y_0$) \{$g=2$; {\bf move} $c_3$ to the right ($y_3$=$y_3+1$);\}

\vspace{1mm}
2.2\; \hspace{4mm}
{\bf else if} ($y_3==y_0$) \{$g=3$; {\bf move} $c_2$ to the left ($y_2$=$y_2-1$);\} 

\vspace{1mm}
2.3\; \hspace{12mm}
{\bf else} \{{\bf move} $c_3$ to the right; {\bf move} $c_2$ to the left;\}


\vspace{1mm}
 2.4\; \hspace{4mm} {\bf move} the robber in any way to try to escape the guard;
 
 \vspace{2mm}
3.\;\; {\bf until} $g\neq 0$;

\vspace{2mm}
4.\;\; {\bf if} ($g==2$) \;$h=3$ \;{\bf else} \;$h=2$\,; \hspace{6mm} // now $c_g$ is the guard 

\vspace{2mm}
5.\;\; {\bf repeat} \hspace{6mm} // cop $c_h$ reaches the initial chasing position 

\vspace{1mm}
5.1.\; \hspace{4mm}{\bf move} $c_g$ horizontally to follow $r$; 

5.2.\; \hspace{4mm}{\bf move} $c_h$ horizontally towards column $\lceil\frac{n}{2} \rceil$;

5.3.\; \hspace{4mm}{\bf move} the robber in any way;

\vspace{2mm}
6.\;\; {\bf until} $c_{h}$ reaches column $ \lceil\frac{n}{2} \rceil$;

\vspace{2mm}
\end{minipage}
}
\end{center}


\item For the CHASE phase all the considerations made for SGRID apply. In particular the shadow cones $\gamma_1,\gamma_2, \gamma_3$ lie below the cops $c_1,c_2,c_3$. 
 As before the robber $r$ must start on the edge of a cone to delay the capture as much as possible, but now the best position for it is  not below row $\lfloor\frac{m}{2}\rfloor$ (Figure \ref{fig:tgrid}), 
otherwise $c_1$ and $c_h$ would chase it ``from the bottom'' of the torus.

\item When $c_1$ and $c_h$ have established a pre-siege, $r$ must move down. The novelty here is that $c_g$ moves towards $r$ in step 1.6, reducing its distance from $r$ hence the number of rounds for the capture.

\end{itemize}

Computing a lower and an upper bound to the capture time of algorithm TGRID implies using floor and ceilings approximations depending on the parity of $m$ and $n$, and on the divisibility of $n$ by 3.
The results are reported in the following Theorem \ref{teo:tgrid}.


\begin{center}
\fbox{
\begin{minipage}{12.2cm}
\small

\vspace{2mm}
{\bf phase} CHASE

\vspace{2mm}
1.\;\; {\bf repeat} \hspace{4mm}//\hspace{1mm} chasing $r$ with cops $c_1$ and $c_h$, while $c_g$ is the guard 


\vspace{1mm}
1.1\; \hspace{4mm}
{\bf if} ($r$ is outside $\gamma_1$ and $ \gamma_h$)   
{\bf move} $c_1$ and $c_h$ horizontally towards $r$

\vspace{1mm}
1.2\; \hspace{4mm}
{\bf else} 
{\bf if} ($r$ is within $\gamma_1$ and/or within $\gamma_h$)   
{\bf move} $c_1$ and $c_h$ down

\vspace{1mm}
1.3\; \hspace{13mm}{\bf else} 
{\bf if} ($r$ is on an edge of $\gamma_1$ (resp. $\gamma_h$) 

\hspace{34mm}{\bf and} outside $\gamma_h$ (rep. $\gamma_1$))  

\hspace{34mm}{\bf move} $c_h$ (resp. $c_1$) horizontally towards that edge 

\vspace{1mm}
1.4\; \hspace{22mm}{\bf else} 
{\bf if} ($r$ is on an edge of $\gamma_1$ and on an edge of $\gamma_h$)  

\hspace{41mm}\{{\bf if} (the cops are in different rows)

\hspace{50mm}{\bf move} down the cop in the highest row

\hspace{43mm}{\bf else} 
{\bf move} down one of the cops\}; 

\vspace{1mm}
1.5\; \hspace{4mm}
{\bf if} ($c_g$ and $r$ are in different columns)   
{\bf move} $c_g$ to the column of $r$

\vspace{1mm}
1.6\; \hspace{3.5mm}
{\bf else} 
{\bf if} ($c_1,c_h$ build a pre-siege)   

\hspace{19mm}{\bf move} $c_g$ from its row $z$ to row $(z-1)$ mod $m$;


\vspace{1mm}
1.7\; \hspace{3.5mm}{\bf move} the robber in any way to try to escape from the cones\,;

\vspace{2mm}
2.\;\; {\bf until} the robber makes its last move inside a siege;

// \hspace{1.5mm}the siege is established with the robber adjacent to $c_g$

\vspace{2mm}
3.\;\; {\bf capture} the robber;

\vspace{2mm}
\end{minipage}
}
\end{center}

\vspace{3mm}

\begin{teo}
\label{teo:tgrid}
In a torus $T_{m,n}$  three cops can capture the robber in time $t$ such that:

\vspace{1mm}
{\em (i)}\; $\frac{2n}{3}+\frac{5m}{4}-\frac{9}{2} \leq t \leq\frac{2n}{3}+ \frac{5m}{4}-\frac{25}{12}$, \;for $m\leq\lceil\frac{n}{2}\rceil$;
 
\vspace{1mm}
{\em (ii)}\, $ \frac{25n}{24}+ \frac{m}{2}-\frac{9}{2} \leq t \leq \frac{25n}{24}+ \frac{m}{2}-\frac{17}{8}$, \;for $\lceil\frac{n}{2}\rceil< m \leq n$.

\vspace{2mm}
\noindent {\bf Proof.} {\em Use algorithm TGRID.}

\noindent Capture. {\em In algorithm GUARD the guard is established by $c_g$ with an obvious procedure, and $c_h$ is brought to column $\lceil\frac{n}{2}\rceil$ to start the chase with $c_1$.
Depending on the position of the robber, either one or both cops move closer to it in each round until a pre-siege is inevitably built around $r$, which is then pushed towards $c_g$ to end in a siege, and then is captured.
}

\vspace{1mm}
\noindent Evaluation of $t$. {\em The algorithm requires three consecutive times $t_1,t_2,t_3$, respectively needed for the guard phase, the construction of a pre-siege, and the construction of a siege, plus an additional round for final capture. $t_1$ is the time to bring $c_h$ to column $\lceil\frac{n}{2}\rceil$, that is at most
$t_1 = \lceil\frac{2n}{3} \rceil -\lceil\frac{n}{2}\rceil$ for $h=2$. The values of $t_2$ and $t_3$ depend on the value of $m$.

(i) Let $m\leq\lceil\frac{n}{2}\rceil$. Starting in row $\lfloor\frac{m}{2}\rfloor$, in the lowest vertex on the edge of a cone and not inside the other cone, the robber forces the cops to apply steps 1.3 and 1.4 as many times as possible until a pre-siege is established in that row while $c_3$ is moved to row $m-1$, and the robber makes a step down to row $\lfloor\frac{m}{2}\rfloor+1$.
As shown in the proof of Theorem \ref{teo:sgrid}, this requires $\lceil\frac{n}{2}\rceil-2$ steps 1.3 plus $2\lfloor\frac{m}{2}\rfloor-1$ steps 1.4, then we have $t_2 = \lceil\frac{n}{2}\rceil+2\lfloor\frac{m}{2}\rfloor-3$. 
 At this point there is a gap of $\lambda$ cop-free rows between the robber $r$ and $c_3$, with $\lambda = m-1-(\lfloor\frac{m}{2}\rfloor+1)-1 = \lceil\frac{m}{2}\rceil-3$. 
 From now on $c_1$ and $c_2$ move down pushing $r$ down, and $c_3$ moves up, until $\lambda$ becomes equal to zero or to one and a siege is established. Since at each round the value of $\lambda$ decreases by two we have $t_3= \lceil(\lceil\frac{m}{2}\rceil-3)/2\rceil=\lceil\frac{m-6}{4}\rceil$. We have $t=t_1+t_2+t_3+1$. With easy approximations to substitute floor and ceiling operators we obtain the bounds specified in the theorem.

(ii) Let $\lceil\frac{n}{2}\rceil< m \leq n$. Even in this case the robber must start on the lowest possible row $j$ on the edge of a cone and not inside the other cone to force the cops to apply steps 1.3 and 1.4 as many times as possible, but the value of $j$ is smaller than in case (i) because the cones have a larger intersection. We have $j=\lceil\frac{n}{4}\rceil$ for $\lceil\frac{n}{2}\rceil$ even, or $j=\lfloor\frac{n}{4}\rfloor$ for $\lceil\frac{n}{2}\rceil$ odd (note that now $j$ does not depend on $m$). 
So $\lceil\frac{n}{2}\rceil-2$ steps 1.3 plus $\lceil\frac{n}{2}\rceil-1$ steps 1.4 are required in the first case, that is $t_2 =2\lceil\frac{n}{2}\rceil-3$ for $\lceil\frac{n}{2}\rceil$ even; 
or $\lceil\frac{n}{2}\rceil-2$ steps 1.3 plus $\lceil\frac{n}{2}\rceil-2$ steps 1.4 are required in the second case, that is $t_2 =2\lceil\frac{n}{2}\rceil-4$ for $\lceil\frac{n}{2}\rceil$ odd. 
The robber is now pushed down along a gap of $\lambda$ cop-free rows, with $\lambda = m-1-(\lceil\frac{n}{4}\rceil+1)-1 = m-\lceil\frac{n}{4}\rceil-3$ for $\lceil\frac{n}{2}\rceil$ even, or $\lambda = m-\lfloor\frac{n}{4}\rfloor-3$ for $\lceil\frac{n}{2}\rceil$ odd, and the siege is reached in $t_3=\lceil\frac{\lambda}{2}\rceil$ rounds.
Also here we have $t=t_1+t_2+t_3+1$, and with easy approximations we obtain the lower and upper bounds specified in the theorem, respectively computed for $\lceil\frac{n}{2}\rceil$ odd and $\lceil\frac{n}{2}\rceil$ even.
}
\hfill $\Box$
\end{teo}


For $T_{7,15}$ of Figure \ref{fig:tgrid}, case (i) of Theorem \ref{teo:tgrid} applies and we have $11.75 <t< 16.67$, that is $12<t<16$ since $t$ must be an integer. Computing $t$ without approximation, using the exact values shown in the proof of the theorem, we have $t_1=2, t_2=11, t_3=1$ hence $t=15$.

We can establish a lower bound on the capture time in a torus identical to the one of Lemma \ref{lem:semitori}, namely:

\begin{lemma}\label{lem:tori}
The capture time in a torus $T_{m,n}$ admits a lower bound $t_{LT}=\lfloor\frac{n}{2}\rfloor+\lfloor\frac{m}{2}\rfloor-2$.

\vspace{2mm}
\noindent {\bf Proof.} {\em For any vertex $u$ of a torus there is a vertex $w$ in a 4-loop whose distance from $u$ is exactly 
$\lfloor\frac{n}{2}\rfloor+\lfloor\frac{m}{2}\rfloor$. If cop $c_h$ is initially placed in $u$, then $r$ can be placed in $w$, and we have from Lemma \ref{lem:siege3}: $t_{LT}=d_h-\lfloor\frac{e}{2}\rfloor =\lfloor\frac{n}{2}\rfloor+\lfloor\frac{m}{2}\rfloor-2$.
}
\hfill $\Box$
\end{lemma}

Letting $t_{UT}$ be the value of $t$ given in Theorem \ref{teo:tgrid} we immediately have:

\begin{cor}
\label{cor:tLB}
In a torus $T_{m,n}$ the ratio $t_{UT}/t_{LT}\rightarrow 4/3$ for $n/m\rightarrow\infty$ and $t_{UT}/t_{LT}\rightarrow \;\sim 37/24$ for $n/m\rightarrow 1$.


\end{cor}

It is worth noting that letting $n<m$, new values for $t_{UT}$ are simply built from the ones of Theorem \ref{teo:tgrid} exchanging $n$ with $m$, while the lower bound $l_{LT}$ of Lemma \ref{lem:tori} holds unchanged. 
So the first statement of Corollary \ref{cor:tLB} is rephrased as: $t_{UT}/t_{LT}\rightarrow 4/3$ for $m/n\rightarrow\infty$. Comparing all these results with the ones found in Theorem \ref{teo:sgrid} and Corollary \ref{cor:sLB} for semi-tori we see that both algorithms improve performance for increasing difference of the grid dimensions. The reason why algorithm SGRID performs comparatively better than TGRID depends in the latter on guard phase  and on the necessity of pushing the robber towards the guard that is at a larger distance than the border of a semi-torus.

\section{Using larger teams of cops}
\label{sec:k-cops}

The cops and robber problem is traditionally focused on studying the minimum number of cops needed for capturing a robber in a given family of graphs, and on the algorithms to successfully attain the capture.   
Let us now take a new approach, discussing how the capture time decreases using an increasing number of cops, and conversely which is the minimum number of cops needed to attain the capture within a given time. 

This approach has a twofold purpose. On one hand, the possibility of employing the cops immediately in a new chase when they have completed their previous job. For example  
assume that a capture can be done by 2 cops in 8 rounds, and by 4 cops in 3 rounds. If 4 cops are available, 2 robbers can be captured in 8 rounds with two parallel chases with 2 cops each, or in 6 rounds with two sequential chases with 4 cops each. Depending on the requests of the problem the latter approach may be preferred. 
The second purpose is completing a job within a required time when a smaller team of cops cannot meet that deadline.

For this new approach we inherit the concept of {\em speed-up} introduced in parallel processing, where the {\em work} $w_k$ of a process carried out by $k$ agents in time $t_k$  is defined as $w_k=k\cdot t_k$, and the speed-up between the actions of $j$ over $i<j$ agents to catch the robber is defined as $w_i/w_j$. If the algorithms run by the two teams of $i$ and $j$ agents are provably optimal, the speed-up is an important measure of the efficiency of parallelism. Referring to the cops and robber problem, the speed-up is a measure of the gain obtained using an increasing number of cops with the best available algorithms.
In this paper we obviously direct our investigation to two-dimensional grids, semi-tori, and tori.


\subsection{$k$ cops on a grid}
\label{sub:k on grid}

Let us consider the case of $k>2$ cops on a grid $G_{m,n}$, with $m\geq 4, n\geq 4$ to avoid trivial cases. W.l.o.g let $m\leq n$. A new algorithm GRID-K can be designed as an extension of algorithm GRID, taking $k$ even. The structure of GRID-K is given below, limited to its main lines for brevity. Still this formulation is sufficient for computing the capture time. 

\begin{figure}[h]
\begin{center}
\includegraphics[scale=0.7]{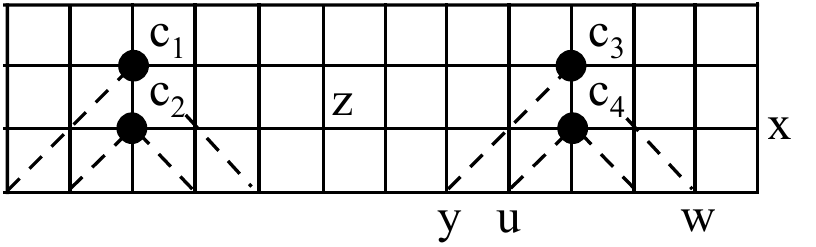}
\end{center}
\caption {\small Two pairs of cops in $G_{4,13}$. \label{fig:pairs}}
\end{figure}

The cops $c_1,\dots,c_k$ are placed in $h$ pairs of adjacent cops, $k=2h$ with $h>1$. The cops of each pair are placed in rows $\lfloor\frac{m}{2} \rfloor-1$ and $\lfloor\frac{m}{2}\rfloor$, and the pairs are almost equally spaced, with 
$\lceil \frac{n-h}{h} \rceil$ and $\lfloor \frac{n-h}{h} \rfloor$ cop-free columns between them, except for the leftmost and the rightmost groups of columns of almost equal sizes whose sum is again $\lceil \frac{n-h}{h} \rceil$ or $\lfloor \frac{n-h}{h} \rfloor$, for example see Figure \ref{fig:pairs} for $k=4$. 


In algorithm GRID-K the robber may be captured on a left or on a right corner of the grid by the leftmost or by the rightmost pair of cops; or it may be captured on the top or on the bottom border by two cops, one from each pair, in a vertex between the two pairs. We have:

\begin{teo}
\label{teo:kgrid}
In a grid $G_{m,n}$, $k=2h$ cops, with $h>1$, can capture the robber in $t_k=\lceil\frac{n-h}{2h}\rceil + \lceil\frac{m-2}{2}\rceil$ rounds.

\vspace{1mm}
\noindent {\bf Proof.} {\em Use algorithm GRID-K.
If the robber is chased by repetitions of steps 1 and 2 it is captured in a corner in $\lceil\frac{n-h}{2h}\rceil + \lceil\frac{m-2}{2}\rceil$ rounds as in algorithm GRID. 
If the robber is chased by repetitions of steps 1 and 3, it is pushed to the border in an almost central vertex between the pairs and is captured there, again in $\lceil\frac{n-h}{2h}\rceil + \lceil\frac{m-2}{2}\rceil$ rounds.}
\hfill $\Box$
\end{teo}


\begin{center}
\fbox{
\begin{minipage}{12.3cm}
\small{
\vspace{1mm}
\noindent{\bf ALGORITHM GRID-K  (SCHEMATIC)}

\vspace{2mm}

Let the cones lay below the cops. 

\begin{enumerate}

\item If the robber $r$ is in both cones of a pair (vertex $u$ of Figure \ref{fig:pairs}), all the cops move vertically towards $r$

\item If $r$ is in a column at the right (resp. left) of the rightmost (resp. leftmost) pair of cops and is not within a cone of the pair (vertices $x$,w of the Figure), $r$ is captured in a corner as in algorithm GRID by repetitions of steps 1 and 2. 

\item If $r$ is in a column between two pairs of cops and not within a cone (vertices $y,z$ of the figure), both pairs of cops move horizontally towards $r$ until it ends in a pair of cones. Then steps 1 and 3 are repeated until $r$ is pushed in a siege on the border  with the concurrence of both pairs of cops.

\vspace{2mm}

\end{enumerate}
}
\end{minipage}
}
\end{center}

For example in $G_{4,13}$ of Figure \ref{fig:pairs} we have $t_4=\lceil\frac{13-2}{4}\rceil + \lceil\frac{4-2}{2}\rceil=4$. The longest capture takes place in the rightmost corner, or in the border between the two pairs of cops. Note that, if computed with $h=1$, the result of Theorem \ref{teo:kgrid} does not coincide with the one of Theorem \ref{teo:grid} for $m$ odd and $n$ even.

\vspace{1mm}
We now compute the minimum number $k$ of cops needed to attain the capture within a given time $t^*$ using algorithm GRID-K, that is the best algorithm known for this problem. From Theorem \ref{teo:kgrid} we have $t_k\geq\frac{n-h}{2h} + \frac{m-2}{2}$ and we easily derive:

\vspace{2mm}
 \hspace{5mm}$k\geq \frac{2n}{2t^* -m+3}\,,$\;\; valid for $G_{m,n}$.\hfill(1)
 
 \vspace{2mm}
 
In the example of Figure \ref{fig:pairs} we have seen that 4 cops capture the robber in 4 rounds. If we  wish to attain the capture in $t^*=3$ rounds we must employ $k\geq \frac{26}{6-4 +3}=5.2$ cops, that is 3 pairs of cops are needed.

\vspace{1mm}
The speed-up for $k=2h$ cops versus 2 cops is given by: 

\vspace{2mm}
\hspace{5mm}$\frac{w_2}{w_k}=2(\lfloor\frac{m+n}{2}\rfloor -1)/2h(\lceil\frac{n-h}{2h}\rceil + \lceil\frac{m-2}{2}\rceil)$.

\vspace{2mm}

\noindent For example for a grid $G_{4,18}$ we have $t=10$ with $k=2$, hence $w_2=20$. Applying algorithm GRID-K with $k=4$ we have $t_4=5$ and $w_4=20$, so the speed-up is one in this case.

\subsection{$k$ cops on a semi-torus}
\label{sub:k on sgrid}

Let us now consider the case of $k>2$ cops on a semi-torus $S_{m,n}$, with $m\geq 3, n\geq 2k$ to avoid trivial cases. A new algorithm SGRID-K, whose main lines are given below, can be built as an immediate extension of algorithm SGRID. This simplified formulation is however sufficient for computing the capture time. 


\begin{center}
\fbox{
\begin{minipage}{12.3cm}
\small{
\vspace{1mm}
\noindent{\bf ALGORITHM SGRID-K  (SCHEMATIC)}

\vspace{2mm}

Let the chase take place in the cones below the cops. 

\begin{enumerate}

\item Until the robber $r$ is within one or more cones, all the cops move down vertically. This eventually brings $r$ on the edge of a cone.

\item Until $r$ is in the gap between two consecutive cops and outside of their cones, the two cops move horizontally towards $r$. This eventually brings $r$ on the edge of a cone.

\item If $r$ is on the edge of a cone, and therefore between two consecutive cops, it is captured by these two cops with algorithm SGRID.
\vspace{2mm}

\end{enumerate}
}
\end{minipage}
}
\end{center}

As for algorithm SGRID, the cops $c_1,\dots,c_k$ are placed in row $\lfloor\frac{m-1}{2} \rfloor$, with $c_1$ in column 0 and the others almost equally spaced along the row, with 
a gap between two consecutive cops of $\lceil\frac{n-k}{k}\rceil$ or $\lfloor\frac{n-k}{k}\rfloor$ cop-free columns according to the value of $n$. 
In the longest chase of algorithm SGRID-K the robber is captured by two cops separated by a larger gap.

Using algorithm SGRID-K, and considering the column gap $\lceil\frac{n-k}{k}\rceil$ instead of $\lceil\frac{n-2}{2}\rceil$ between the leftmost cops in the proof of Theorem \ref{teo:sgrid}, we have with straightforward computation:

\begin{teo}
\label{teo:ksgrid}
In a semi-torus $S_{m,n}$  $k\geq 2$ cops can capture the robber in time:

\vspace{1mm}
{\em (i)}\;\, $t_k=\lceil\frac{n}{k} \rceil+ 2\lfloor\frac{m}{2}\rfloor -2$,\; for $\lfloor\frac{m}{2}\rfloor \leq \lceil\frac{n-k}{2k} \rceil $;
 
\vspace{1mm}
{\em (ii)}\, $t_k=\lceil\frac{n}{k} \rceil+  \lceil\frac{n-k}{2k} \rceil + \lfloor\frac{m}{2}\rfloor -2$,\; for $\lfloor\frac{m}{2}\rfloor > \lceil\frac{n-k}{2k} \rceil $.

\end{teo}

\vspace{1mm}
Theorem \ref{teo:ksgrid} is an immediate extension of Theorem \ref{teo:sgrid} and coincides with it for $k=2$. We now compute the minimum number $k$ of cops needed to attain the capture within a given time $t^*$ using algorithm SGRID-K. From Theorem \ref{teo:ksgrid} we have: 

\vspace{3mm}
\noindent Case (i) \;\;$t_k\geq \frac{n}{k}+m-2$, \;hence $k\geq\frac{n}{t^* -m+2}$\,,\;\; for $m$ even;\hfill(2.1)

\vspace{1mm}
\hspace{1.1cm}$t_k\geq \frac{n}{k}+m-3$, \;hence $k\geq\frac{n}{t^* -m+3}$\,,\;\; for $m$ odd.\hfill(2.2)

\vspace{2mm}
\noindent Case (ii) \;\,$t_k\geq \frac{n}{k}+\frac{n-k}{2k}+\frac{m}{2}-2$, \;hence $k\geq\frac{3n}{2t^* -m+5}$\,,\;\; for $m$ even;\hfill(2.3)

\vspace{1mm}
\hspace{1.1cm}\,$t_k\geq \frac{n}{k}+\frac{n-k}{2k}+\frac{m}{2}-3$, \;hence $k\geq\frac{3n}{2t^* -m+7}$\,,\;\; for $m$ odd.\hfill(2.4)

\vspace{3mm}
\noindent In relations (2.1) to (2.4) note that, for a given $m$ the desired time $t^*$ must be large enough to make the denominator greater than zero.

\begin{figure}[h]
\begin{center}
\includegraphics[scale=0.7]{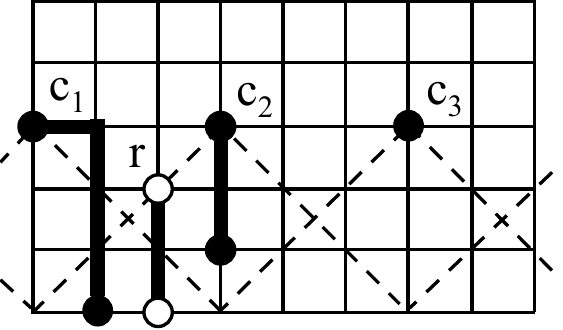}
\end{center}
\caption {\small Movements of three cops and the robber in $S_{6,9}$, up to a siege: the last two moves of $c_1$ are done concurrently with the moves of $c_2$. \label{fig:ksgrid}}
\end{figure}

As an example of speed-up consider the semi-torus $S_{6,9}$ in Figure \ref{fig:ksgrid}. For $k=2$ (Figure \ref{fig:2sgrid}) we have already found $t=8$ hence $w_2=16$. For $k=3$, case (ii) of Theorem \ref{teo:ksgrid} applies and we have $t_3=\lceil\frac{n}{k} \rceil+  \lceil\frac{n-k}{2k} \rceil + \lfloor\frac{m}{2}\rfloor -2=5$, hence $w_3=15$ and $\frac{w_2}{w_3}>1$. This is a case of super-linear speed-up computed with the best available algorithms for semi-tori. Recall that the speed-up may be different and clearly more significant using provably optimal algorithms if they were known.



\subsection{$k$ cops on a torus}
\label{sub:k on tgrid}

Let us now consider $k$ cops working on a torus $T_{m,n}$ with $k\geq 4$. W.l.o.g. let $n\geq m$, and let $m\geq 6, n\geq 2k$ to avoid trivial cases. As before a schematic formulation of TGRID-K, given as an immediate extension of TGRID, is sufficient for computing the capture time. 

The $k$ cops are placed in row 0 in the order $c_1,c_k,c_2,c_3,\dots,c_{k-1}$, with the first in column 0 and the others almost equally spaced along the row, with 
a gap between two consecutive cops of $\lceil\frac{n-k}{k}\rceil$ or $\lfloor\frac{n-k}{k}\rfloor$ cop-free columns according to the value of $n$. Assume that the larger gaps occur between the cops at the beginning of the sequence, so $c_k$ and $c_2$ respectively start in columns $\lceil\frac{n}{k}\rceil$ and $2\lceil\frac{n}{k}\rceil$.
W.l.o.g. assume that cop $c_k$ will be the guard and the longest chase will be done by $c_1$ and $c_2$.


\begin{center}
\fbox{
\begin{minipage}{12.3cm}
\small{
\vspace{1mm}
\noindent{\bf ALGORITHM TGRID-K  (SCHEMATIC)}

\vspace{2mm}

Let the robber start in the gap between $c_k$ and $c_2$.

\begin{enumerate}

\item GUARD PHASE. 

$c_k$ moves rightwards and $c_2,\dots,c_{k-1}$ move leftwards, concurrently in row 0, until they  reach their proper positions for the chase. $c_k$ eventually becomes the guard and the phase ends when $c_2$ reaches column $\lceil\frac{n}{k-1}\rceil$.

\item CHASE PHASE. 

2.1 While the robber $r$ is within one or more cones, all the cops except $c_k$ move down vertically. This eventually brings $r$ on the edge of a cone. If needed, $c_k$ moves horizontally to stay in the same column of $r$.

\vspace{1mm}
2.2 While $r$ is in the gap between two consecutive cops (assume that they are $c_1$ and $c_2$ for the longest chase), or on the edge of one or both cones, it is captured by these two cops as in the CHASE phase of algorithm TGRID run by them together with the guard $c_k$.

\vspace{2mm}
\end{enumerate}
}
\end{minipage}
}
\end{center}

In the guard phase of algorithm TGRID-K the guard is established by $c_k$ and the cops $c_1,\dots,c_{k-1}$ are brought to almost equally spaced positions in row 0 (the new gaps will be $\lceil\frac{n-(k-1)}{k-1}\rceil$ or $\lfloor\frac{n-(k-1)}{k-1}\rfloor$) to be prepared for chasing the robber (which, in the longest chase, will be captured by $c_1$ and $c_2$). For this purpose $c_1,\dots,c_{k-1}$ move together rightwards for the needed number of steps, depending on the sizes of the gaps between the cops. 
In any case $c_2$ is placed in column $\lceil\frac{n}{k-1}\rceil$ with  $2\lceil\frac{n}{k}\rceil-\lceil\frac{n}{k-1}\rceil$ moves, and no other cop makes more moves in this phase of the algorithm.
In the chase phase of the algorithm, first the robber is confined in a set of columns between two cops (say $c_1$ and $c_2$), then is chased as in TGRID in this narrower section of the torus. 

For torus $T_{7,15}$ with $k=4$, the initial positions of the cops and the robber, and their evolution according to algorithm TGRID-K, are indicated in Figure \ref{fig:ktgrid}.
The analysis of TGRID-K is an extension of the one of TGRID. We have:


\begin{figure}[h]
\begin{center}
\includegraphics[scale=0.7]{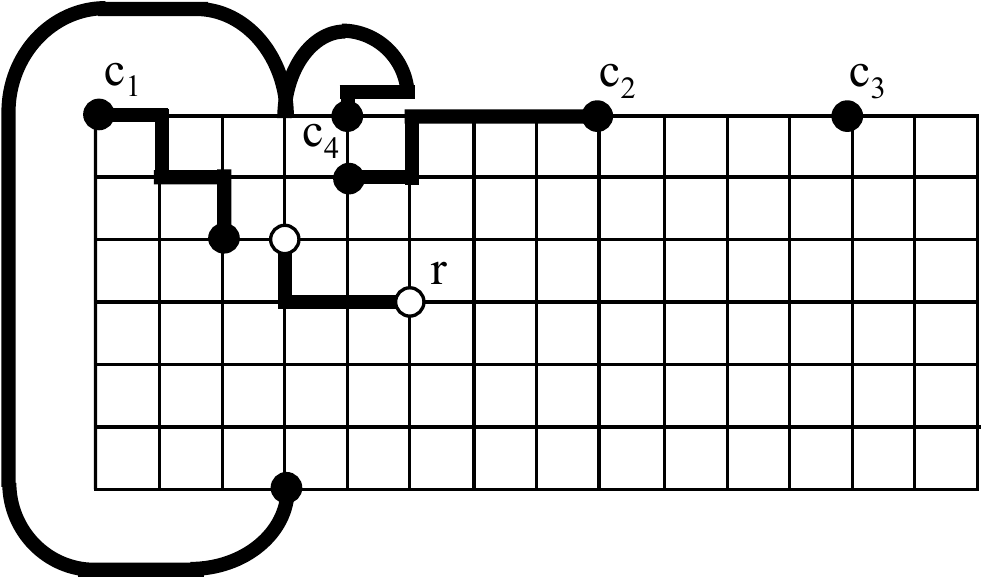}
\end{center}
\caption {\small Chase with four cops in $T_{7,15}$ up to a pre-siege. The first three moves of $c_4,c_2$, and $r$ take place in the GUARD phase. Compare the moves with the ones for $k=3$ reported in figure \ref{fig:tgrid}.\label{fig:ktgrid}}
\end{figure}


\begin{teo}
\label{teo:ktgrid}
In a torus $T_{m,n}$, $k>3$ cops can capture the robber in time $t_k$ such that:

\vspace{1mm}
{\em (i)}\; $\frac{2n}{k}+\frac{5m}{4}-\frac{9}{2} \leq t_k \leq \frac{2n}{k}+ \frac{5m}{4}+\frac{k-1}{k}-\frac{11}{4}$, \;for $m\leq\lceil\frac{n}{k-1}\rceil$;
 
\vspace{1mm}
{\em (ii)}\, $ \frac{2n}{k}+ \frac{3n}{4(k-1)}+\frac{m}{2}-\frac{9}{2} \leq t _k< \frac{2n}{k}+ \frac{3n}{4(k-1)}+\frac{m}{2}-\frac{1}{2}$, \;for $\lceil\frac{n}{k-1}\rceil< m \leq n$.

\vspace{2mm}
\noindent {\bf Proof}. {\em Use algorithm TGRID-K, and refer to the proof of Theorem \ref{teo:tgrid} for comparison. Attaining the capture is obvious.  The guard time is  $t_1=  \lceil\frac{2n}{k}\rceil-\lceil\frac{n}{k-1}\rceil$. 

\vspace{1mm}
\noindent In case (i), the time to establish a pre-siege  
is $t_2=\lceil\frac{n}{k-1}\rceil+2\lfloor\frac{m}{2}\rfloor-3$, and the following time to establish a siege is $t_3=\lceil\frac{m-6}{4}\rceil$. With proper approximations of the ceiling and floor functions the capture time $t_k=t_1+t_2+t_3+1$ can be bounded as in the statement of the theorem. 

\vspace{1mm}
\noindent In case (ii) the time $t_2$ to establish a pre-siege depends on the parity of $\lceil\frac{n}{k-1}\rceil$. The robber is placed in row $\lceil\frac{n}{2(k-1)}\rceil$ for $\lceil\frac{n}{k-1}\rceil$ even, or in row $\lfloor\frac{n}{2(k-1)}\rfloor$ for $\lceil\frac{n}{k-1}\rceil$ odd. Refer to the CHASE phase of algorithm TGRID used inside TGRID-K. In the first case $\lceil\frac{n}{k-1}\rceil-2$ steps 1.3 plus $\lceil\frac{n}{k-1}\rceil-1$ steps 1.4 are required, that is $t_2=2\lceil\frac{n}{k-1}\rceil-3$ for $\lceil\frac{n}{k-1}\rceil$ even. In the second case $\lceil\frac{n}{k-1}\rceil-2$ steps 1.3 plus $\lceil\frac{n}{k-1}\rceil-2$ steps 1.4 are required, that is $t_2=2\lceil\frac{n}{k-1}\rceil-4$ for $\lceil\frac{n}{k-1}\rceil$ odd.
The following time to establish a siege is $t_3=\lceil\frac{\lambda}{2}\rceil$, with $\lambda=m-\lceil\frac{n}{2(k-1)}\rceil-3$ for $\lceil\frac{n}{k-1}\rceil$ even, and $\lambda=m-\lfloor\frac{n}{2(k-1)}\rfloor-3$ for $\lceil\frac{n}{k-1}\rceil$ odd.
With proper approximations of the ceiling and floor functions the capture time $t_k=t_1+t_2+t_3+1$ can be bounded as in the statement of the theorem.}
\hfill $\Box$

\end{teo}

Note that the bounds for $t_k$ given in Theorem \ref{teo:ktgrid} coincide with the ones of Theorem \ref{teo:tgrid} for $k=3$, except for the upper bound of case (ii) that has now been evaluated with a stronger approximation (thus yielding a $<$ sign instead of the stricter $\leq$ sign) to avoid a complicated formula.
In the torus $T_{7,15}$ of Figure \ref{fig:ktgrid}, using $k=4$ cops and applying the exact values of the numbers of steps reported in the proof of Theorem \ref{teo:ktgrid} case (ii) with $\lceil\frac{n}{k-1}\rceil=5$ odd we have: $t_4=3+6+1+1=11$.

The minimum number $k$ of cops needed to attain the capture within a given time $t^*$ using algorithm TGRID-K is derived in the two cases of Theorem \ref{teo:ktgrid} with some further approximations. We have::

\vspace{2mm}
\noindent Case (i) \;\;$\frac{8n}{4t^* -5m+18}\leq k<\frac{8n}{4t^* -5m+7}$\,,\;\; for $m\leq\lceil\frac{n}{k-1}\rceil$; \hfill(3.1)

\vspace{2mm}
\noindent Case (ii) \;\,$\frac{11n}{4t^* -2m+18}<k<\frac{11n}{4t^* -2m+2}+1$\,,\;\;for $\lceil\frac{n}{k-1}\rceil< m \leq n$.  \hfill(3.2)

\vspace{3mm}

For torus $T_{7,15}$ of Figure \ref{fig:ktgrid} case (ii) applies for any $k>3$. Imposing a capture time of at most $t^*=12$, from relation (3.2) we have $3.17<k<5.58$, that is the required number of cops is between 4 and 5.
In fact we have already seen that 3 cops require 15 rounds and 4 cops require 11 rounds. This also implies that 
$w_3=45$ and $w_4=44$, hence a slightly super-linear speed-up occurs.


\section{Concluding remarks}

In this work we have extended the well known cops and robber problem in two-dimensional grids to semi toroidal and fully toroidal grids. We have introduced the concepts of {\em siege} around the robber and of {\em shadow-cone} of a cop to reconstruct known results on grids, and we have used these tools for studying the new chase on toroidal grids, giving efficient algorithms for different instances of the problem. 
Although we have not been able to prove that our algorithms are optimal for toroidal grids in relation to the capture time, we have shown that their behaviour tends to be optimal if the ratio between the numbers of rows and columns becomes unbalanced.

We have then discussed the effect of using an arbitrary (i.e. non necessarily minimal) number of cops, studying new algorithms for this case and computing the minimum number of cops needed if the capture time is fixed. For this purpose we have
inherited the concept of {\em work} from parallel processing, for computing the speed-up obtained if the number of cops increases, as an indication of the effect of using a large number of cops.  In the realm of our algorithms we have shown that even super-linear speed-up may occur.

Two main extensions of our work are now in order. One is improving the algorithms for semi-tori and for tori, and/or determining higher lower bounds on the capture time, with the final goal of obtaining optimal algorithms in relation to the capture time once the number of cops is fixed. The second is extending our study to multi-dimensional grids and tori. In addition it may be worth extending our approach to the capture on different graphs, and possibly to other classes of problems dealing with mobile agents. The present work is to be seen as a first step in this direction.


\begin{thebibliography}{10}

 \bibitem{AF84}
M. Aigner and M. Fromme.
 \newblock A game of cops and robbers.
 \newblock {\em Discrete  Applied Mathematics} 8, 1-12, 1984.

 \bibitem{A04}
B. Alspach.
 \newblock Searching and sweeping graphs: a brief survey.
 \newblock {\em Le Matematiche} 59 (I-II), 5-37, 2004.

\bibitem{B+10}
S. Bhattacharya, G. Paul and S. Sanyal.
 \newblock A cops and robber game in multidimensional grids.
 \newblock {\em Discrete  Applied Mathematics} 158, 1745-1751, 2010.

\bibitem{B+05}
S. Bhattacharya, A. Banerjee and S. Badyopadhay.
 \newblock CORBA-based analysis of multi-agent behavior.
 \newblock {\em Journal of Computer Science and Technology} 20 (1), 118-124, 2005.

\bibitem{B+06}
L. Blin, P. Fraignaud, N. Nisse, and S. Vial.
 \newblock Distributed chasing of network intruders.
 \newblock {\em Theoretical Computer Science} 399, 12-37, 2008.
 
 \bibitem{B+09}
A. Bonato, P. Golovach, G. Hahn, and J. Kratochvil.
 \newblock The capture time of a graph.
\newblock {\em Dicrete Mathematics} 309, 5588-5595, 2009.

\bibitem{BN11}
A. Bonato and R. Nowakovski.
 \newblock {\em The Game of Cops and Robbers on Graphs}.
\newblock American Mathematical Society, 2011.

 \bibitem{C+16}
N. Cohen, M.Hilaire, N.A. Martins, N. Nisse, and S. Perennes.
 \newblock Spy-game on graphs.
 \newblock {\em Proc. 8-th International Conference FUN 2016} DOI 10.4230/LIPIcs.FUN.2016.10

\bibitem{D92}
R. Dawes.
 \newblock  Some pursuit-evasion problems on grids.
 \newblock{\em Information Processing Letters} 43, 241-247, 1992.

 \bibitem{DKSZ08}
A. Dumitrescu, H. Kok, I. Suzuki and P. Zylinski.
 \newblock  Vision based pursuit-evasion on a grid.
 \newblock {\em Proc. 11-th Scandinavian Workshop on Algorithm Theory, SWAT 2008}  LNCS 5124, 45-64, 2008.

 \bibitem{EW08}
J. Ellis and R. Warren.
\newblock Lower bounds on the pathwidth of some grid-like graphs.
 \newblock{\em Discrete Applied Mathematics} 156, 545-555, 2008.

 \bibitem{F+10}
F. Fomin, P. Golovach, J. Kratochvil, N. Nisse and K. Suchan.
 \newblock  Pursuing a fast robber on a graph.
 \newblock {\em Theoretical Computer Science} 411, 1167-1181, 2010

  \bibitem{INS09}
D. Ilcinkas, N. Nisse and D. Soguet.
 \newblock  The cost of monotonicity in distributed graph searching.
 \newblock {\em Distributed Computing} 22(2)  117-127, 2009

  \bibitem{GR95}
 F. Goldstein and E. Reingold.
 \newblock The complexity of pursuing a graph.
 \newblock {\em Theoretical Computer Science} 143, 93-112, 1995

\bibitem{KR90}
R.M. Karp and V. Ramachandran.
\newblock Parallel algorithms for shared memory machines.
\newblock In: J. van Leeuwen (ed) {\em  Handbook of Theoretical Computer Science, Vol. A}.
 \newblock{North Holland, New York, 869-941, 1990}.

\bibitem{K15}
W.B. Kinnersley.
\newblock Cops and Robbers is EXPTIME-complete.
\newblock {\em  Journal of Combinatorial Theory, Series B} 111, 201-220, 2015.
 
 \bibitem{LPP92}
 F. Luccio, L. Pagli, and G. Pucci.
 \newblock Three non Conventional Paradigms of Parallel Computation.
 \newblock In: {\em Parallel Architectures and Their Efficient Use.}
  \newblock LNCS 678, 166-175, 1992.


\bibitem{LP09}
 F. Luccio and L. Pagli.
 \newblock A general approach to toroidal mesh decontamination with local immunity.
 \newblock {\em Proceedings of the 23rd IEEE International  Parallel and Distributed
 Processing Symposium} (IPDPS),  1-8, 2009.

  \bibitem{LP16}
 F. Luccio and L. Pagli.
 \newblock More agents may decrease global work: A case in butterfly decontamination.
\newblock{\em Theoretical Computer Science} 655, 41-57, 2016.

 \bibitem{MM87}
 M. Maamoun and H. Meyniel.
 \newblock On a game of policemen and robber.
 \newblock {\em Discrete Applied Mathematics} 17, 18--44, 1988.

 \bibitem{Me+88}
 N. Megiddo, S. Hakimi, M. Garey, D. Johnson and C. Papadimitriou.
 \newblock The complexity of searching a graph.
 \newblock {\em Journal of the ACM} 35 (1), 307-309, 1987.

 \bibitem{M11}
A. Mehrabian.
 \newblock The capture time of grids.
 \newblock {\em Discrete Mathematics} 311, 102-105, 2011.

 \bibitem{N96}
 S. Neufeld.
 \newblock A pursuit-evasion problem on a grid
 \newblock{\em Information Processing Letters} 58, 5-9, 1996.
 
 \bibitem{NN98}
 S. Neufeld and R. Nowakovsky
 \newblock A game on cops and robbers played on products of graphs.
 \newblock{\em Discrete Mathematics} 186, 253-268, 1998.

 \bibitem{NW83}
R. Nowakowski and P. Winkler.
\newblock Vertex-to-vertex pursuit in a graph.
 \newblock{\em Discrete Mathematics} 43, 253-259, 1983.
 
 \bibitem{PX16}
P. Pisantechakool and X. Tan
\newblock On the capture time of cops and robbers game on a planar graph.
 \newblock In: T-H.H. Chan et al., Eds. {\em Proc. COCOA 2016,  LNCS} 10043, 3-17, 2016.

\bibitem{Q85}
A. Quillot. 
 \newblock  These di $3^{\circ}$ cycle.
 \newblock {\em UniversitŽ de Paris VI}, 131-145, 1978.

\bibitem{SS89}
K. Sugihara and I. Suzuki
\newblock  Optimal Algorithm for a pursuit-evasion problem.
 \newblock{\em SIAM Journal of Discrete Mathematics} 2, 126-143, 1989.




\end{thebibliography}
\end{document}